\input{aipcheck}

\documentclass[
    ,final            
  ]
  {aipproc}

\layoutstyle{6x9}
 \usepackage{epsfig}
 \usepackage{rotating}
 \usepackage{amsmath,amssymb}
 \usepackage{graphicx}
 \usepackage{setspace}
 \usepackage{fancyhdr}
 \usepackage{moreverb}
 \usepackage{rotating}

\def\e6{$E(6)$}
\def\10{$SO(10)$}
\def\21{$SU(2) \otimes U(1) $}
\def\lr{$SU(2)_L \otimes SU(2)_R \otimes U(1)$}
\def\422{$SU(4) \otimes SU(2) \otimes SU(2)$}
\def\321{$SU(3) \otimes SU(2) \otimes U(1)$}
\def\lsim{\raise0.3ex\hbox{$\;<$\kern-0.75em\raise-1.1ex\hbox{$\sim\;$}}}
\def\gsim{\raise0.3ex\hbox{$\;>$\kern-0.75em\raise-1.1ex\hbox{$\sim\;$}}}

\def\meff{\langle m_{\nu} \rangle}
\newcommand{\ed}{\end{document}}
\DeclareMathAlphabet{\mathsc}{OT1}{cmr}{m}{sc}

\newcommand{\nue}{\hbox{$\nu_e$ }} 
\newcommand{\nm}{\hbox{$\nu_\mu$ }} 

\newcommand{\nt}{\hbox{$\nu_\tau$ }} 

\newcommand{\CL}   {C.L.}
\newcommand{\dof}  {d.o.f.}
\newcommand{\gof}{g.o.f.}

\newcommand{\eVq}  {\rm{eV}^2}
\newcommand{\Sol}  {\mathsc{sol}}
\newcommand{\Atm}  {\mathsc{atm}}

\newcommand{\Sbl}  {\mathsc{sbl}}
\newcommand{\Kl}  {\mathsc{KamLAND}}

\newcommand{\Lsnd} {\mathsc{lsnd}}

\newcommand{\Dcq}  {\Delta\chi^2}
\newcommand{\Dms}  {\Delta m^2_\Sol}
\newcommand{\Dma}  {\Delta m^2_\Atm}
\newcommand{\Dml}{\Delta m^2_\Lsnd}
\newcommand{\Eps}  {\varepsilon}
\newcommand{\Epp}  {\varepsilon'}

\newcommand{\snocc}{SNO$_\mathrm{CC}^\mathrm{rate}$ }
\newcommand{\snotot}{SNO$_\mathrm{CC,NC}^\mathrm{SP,DN}$ }

\newcommand{\pnu}[1] {\overset{\smash{\scriptscriptstyle (-)}}{\nu}_{\hskip-3pt #1}}

\def \nbb {$\beta\beta_{0\nu}$ }

\def\meff{\langle m_{\nu} \rangle}

\newcommand{\AddrAHEP}{%
 Instituto de F\'{\i}sica Corpuscular,
  C.S.I.C. -- Universitat de Val{\`e}ncia \\
  Edificio de Institutos de Paterna, Apartado 22085,
  E--46071 Val{\`e}ncia, Spain\\}
\newcommand{\AddrTub}{%
Institut für Theoretische Physik, Auf der Morgenstelle 14\\
Universität T\"ubingen,
D-72076 Tübingen, Germany}

\begin{document}

\title{Neutrino masses twenty--five years later}

\author{J. W. F. Valle}{
  address={\AddrAHEP and \\ \AddrTub }
}

\begin{abstract}
  
  The discovery of neutrino mass marks a turning point in elementary
  particle physics, with important implications for nuclear and
  astroparticle physics.  Here I give a brief update, where I
  summarize the current status of three--neutrino oscillation
  parameters from current solar, atmospheric, reactor and accelerator
  neutrino data, discuss the case for sterile neutrinos and LSND, and
  also the importance of tritium and double beta decay experiments
  probing the absolute scale of neutrino mass.  In this opininated
  look at the present of neutrino physics, I keep an eye in the
  future, and a perspective of the past, taking the oportunity to
  highlight Joe Schechter's pioneering contribution, which I have had
  the fortune to share, as his PhD student back in the early eighties.

\end{abstract}

\maketitle


\section{Introduction}

The basic theoretical setting required for the description of current
neutrino oscillation experiments
\cite{fukuda:1998mi,fukuda:2002pe,ahmad:2002jz,ahmad:2002ka,eguchi:2002dm}
was laid out in the early
eighties~\cite{schechter:1980gr,cheng:1980qt,schechter:1981hw,mohapatra:1981yp}.
This included the two-component quantum description of massive
Majorana neutrinos and the gauge theoretic characterization of the
lepton mixing matrix describing neutrino
oscillations~\cite{schechter:1980gr}. To complete the formulation of
neutrino oscillations necessary to describe current data the other
important ingredient was the formulation of the theory of neutrino
oscillations in matter by Mikheev, Smirnov and
Wolfenstein~\cite{mikheev:1985gs,wolfenstein:1978ue}~\footnote{For
  recent reviews
  see~\cite{pakvasa:2003zv,gonzalez-garcia:2002dz,bilenky:1998rw} and
  references therein.}.  The theoretical origin of neutrino mass
remains as much of a mystery today as it was back in the eighties.
Much of the early effort devoted to the study of neutrino masses was
motivated in part by the idea of unification which introduced the
seesaw mechanism~\cite{gell-mann:1980vs,yanagida:1979}, and by the
(later unconfirmed) hints for neutrino oscillations then seen by
Reines~\cite{reines:1980pc}.  The ba sic dimension--five neutrino mass
operator~\cite{weinberg:1980mh} arises in the context of the \10
unification group, though it was soon realized that the seesaw idea
can be applied to left-right symmetric
theories~\cite{mohapatra:1981yp}, or the simplest effective Standard
Model gauge framework~\cite{schechter:1980gr}.  While the \10 or \lr
formulations of the seesaw have the virtue of relating the small
neutrino mass to the dynamics of parity (gauged B-L) violation, the
effective \21 description is more general and applies to any theory,
for example with ungauged
B-L~\cite{chikashige:1981ui,schechter:1982cv}.  The role of a Higgs
triplet in generating neutrino masses was noted in the early days,
either on its own or as part of the seesaw
mechanism~\cite{schechter:1980gr,cheng:1980qt,mohapatra:1981yp}~\footnote{Such
  triplet contribution to neutrino mass may come from an induced
  vacuum expectation value.}.  The detailed general structure of the
seesaw diagonalizing matrix given in Ref.~\cite{schechter:1982cv}
plays a role in the determination of the baryon asymmetry of the
Universe, within the so-called leptogenesis
scenarios~\cite{kuzmin:1985mm,fukugita:1986hr}.

A radical approach to the seesaw idea is that of neutrino unification,
recently advocated in~\cite{chankowski:2000fp} and \cite{babu:2002dz}
leads naturally to a quasi degenerate neutrino spectrum with important
phenomenological implications.

However, it is worth stressing that the seesaw is only one way of
generating the basic dimension--five neutrino mass
operator~\cite{weinberg:1980mh}, which may arise from physics ``just
around the corner'', such as low energy
supersymmetry~\cite{diaz:2003as,hirsch:2000ef,romao:1999up}.  For
example, schemes with spontaneously broken R
parity~\cite{ross:1985yg,ellis:1985gi,masiero:1990uj} lead effectively
to bilinear R parity violation~\cite{diaz:1998xc}.  Neutrino mixing
angles can be tested at accelerator
experiments~\cite{hirsch:2002ys,porod:2000hv,restrepo:2001me} Other
variants of this idea involve triplet Higgs bosons, such as the model
in~\cite{aristizabal:2003ix}.  Alternative low energy mechanisms for
neutrino mass generation are the models of Babu~\cite{babu:1988ki} and
Zee~\cite{zee:1980ai} and variants thereof.

Much is now known about the structure of the lepton mixing matrix
since the paper in Ref.~\cite{schechter:1980gr} has been written.
First, LEP data imply three light sequential \21 doublet (active)
neutrinos~\cite{hagiwara:2002fs}, $\nu_e$, $\nu_{\mu}$ and
$\nu_{\tau}$.  This still leaves open the possibility of singlet
leptons remaining as light as the electron-volt range, due to some
symmetry~\cite{peltoniemi:1993ec,peltoniemi:1993ss,caldwell:1993kn}.
In this case they might take part in the oscillations as sterile
neutrinos~\cite{giunti:2000vv}, as hinted by the data of
LSND~\cite{aguilar:2001ty}. While this experiment is currently
unconfirmed, a global analysis of all current oscillation experiments
strongly prefer the minimal three light--neutrino
hypothesis~\cite{maltoni:2002ni,maltoni:2002xd}.  The possibility of
symmetric (2+2) schemes is ruled out, because in this case sterile
neutrinos take part in both solar and atmospheric oscillations. On the
other hand the presence of a light sterile neutrino in a (3+1) scheme
is still allowed, since it can be chosen to decouple from solar and
atmospheric oscillations, though strongly disfavoured by
short-baseline experiments.

Data from cosmology, including CMB data from
WMAP~\cite{spergel:2003cb}
\cite{crotty:2003th,elgaroy:2003yh,hannestad:2003xv} and the 2dFGRS
large scale structure surveys~\cite{tegmark:2001jh} lead to further
restrictions, especially on large $\Dml$ values \footnote{Note that it
  is not possible to rule out light sterile states if they do not mix
  significantly with the three sequential active neutrinos.  Though
  heavy isosinglet leptons would not be emitted in weak decays, they
  would lead to an effectively non-unitary lepton mixing matrix
  characterizing the three active neutrinos~\cite{schechter:1980gr}.}.
The three--neutrino lepton mixing matrix is characterized by three
mixing angles: $\theta_{12}$ which describes solar neutrino
oscillations, $\theta_{23}$ which characterizes atmospheric neutrino
oscillations, and $\theta_{13}$ which couples these two analyses. It
can be written as
$$
K = \omega_{12} \omega_{13} \omega_{23}
$$
where $\omega_{ij}$ is a complex rotation in the $ij$ sector.  This
parametrization can be found in \cite{schechter:1980gr}.  The charged
current lepton mixing matrix also contains one Kobayashi-Maskawa--like
leptonic CP phase whose effects are suppressed due to the stringent
limits on $\theta_{13}$ following mainly from reactor
data~\cite{apollonio:1999ae}. Moreover they are suppressed by the
small mass splitting indicated by the solar neutrino data analysis
(see below). This happens since, in the 3-neutrino limit, CP violation
disappears as two neutrinos become degenerate~\cite{schechter:1980bn}.
The effect of this phase in current neutrino oscillation experiments
is negligible.  Future neutrino
factories~\cite{apollonio:2002en,freund:2001ui,albright:2000xi,cervera:2000kp}
aim to be sensitive to such CP violating effects.
There are, in addition, two CP phases associated with the (12) and
(23) sectors, that can not be removed by field redefinition due to the
Majorana nature of neutrinos~\cite{schechter:1980gr}.  However these
non--Kobayasji--Maskawa--like CP phases drop out from conventional
oscillation experiments.  All CP phases are neglected in current
oscillation analyses, which take the matrices $ \omega_{ij}$ as real
rotations.

However these Majorana phases do affect lepton-number-violating
oscillations~\cite{schechter:1981gk}. Unfortunately, these are
strongly suppressed by the small masses of neutrinos and the V-A
nature of the weak interaction.  Massive majorana neutrinos are also
expected to have non-zero transition magnetic
moments~\cite{schechter:1981hw}, sensitive to the Majorana
phases~\cite{schechter:1981hw,kayser:1982br,nieves:1982zt}. However
magnetic moments also vanish in the limit of massless
neutrinos~\cite{pal:1982rm}.  Neutrinoless double beta
decay~\cite{doi:1981yb,wolfenstein:1981rk} holds better chances of
revealing the effects of these extra phases. However nuclear physics
uncertainties currently preclude a realistic way to test
them~\cite{Barger:2002vy}.

\section{Neutrino data analysis}

Neutrino masses have been discovered in atmospheric
neutrinos~\cite{fukuda:1998mi}.  In contrast solar neutrino
experiments could not establish neutrino
oscillations~\cite{barranco:2002te,guzzo:2001mi} without the results
of KamLAND. An analysis of recent solar, atmospheric and reactor
data~\footnote{See Ref.~\cite{maltoni:2002ni} for an extensive list of
  solar and atmospheric neutrino experiments.}  has been given
in~\cite{maltoni:2002ni}~\footnote{For a discussion of other neutrino
  oscillation analyses see Table 1 in ~\cite{pakvasa:2003zv} and Table
  2 in \cite{maltoni:2002ni}}.  This paper presents a generalized
determination of the neutrino oscillation parameters taking into
account that both the solar \nue and the atmospheric \nm may convert
to a mixture of active and sterile neutrinos.  This allows one to
systematically combine solar and atmospheric data with the current
short baseline neutrino oscillation data including the LSND evidence
for oscillations~\cite{maltoni:2002xd}.

Insofar as atmospheric data are concerned, the analysis in
\cite{maltoni:2002ni} used the data given in
Refs.~\cite{fukuda:1998ah,fukuda:1998tw,fukuda:1998ub,fukuda:1999pp}
as well as the most recent atmospheric Super-K
(1489-day)~\cite{shiozawa:2002} and MACRO~\cite{surdo:2002rk} data.

The solar neutrino data include rates for the Homestake chlorine
experiment~\cite{cleveland:1998nv} ($2.56 \pm 0.16 \pm 0.16$~SNU), the
most recent result of the gallium experiments
SAGE~\cite{abdurashitov:1999bv}~($70.8 ~^{+5.3}_{-5.2}
~^{+3.7}_{-3.2}$~SNU) and GALLEX/GNO~\cite{altmann:2000ft} ($70.8 \pm
4.5 \pm 3.8$~SNU), as well as the 1496-days Super-K data
sample~\cite{fukuda:2002pe}. The latter are presented in the form of
44 bins (8 energy bins, 6 of which are further divided into 7 zenith
angle bins).
In addition, we have the latest SNO results in
Refs.~\cite{ahmad:2002jz}, in the form of 34 data bins (17 energy bins
for each day and night period). All in all, $3+44+34=81$ observables.

Moreover, in version 3 of the arXiv the implications of the first
145.1 days of KamLAND data on the determination of the solar neutrino
parameters are also discussed in detail, updating the published
version~\cite{maltoni:2002ni}.

\subsection{Atmospheric + reactor}

The atmospheric plus reactor data can be well described in the
approximation $\Dms\ll\Dma$, taking the electron neutrino as
completely decoupled from atmospheric oscillations, with $\theta_{13}
\to 0$ (for the case $\theta_{13} \neq 0$ see below).

\begin{figure}[bthp] \centering
    \includegraphics[width=0.9\linewidth,height=7cm]{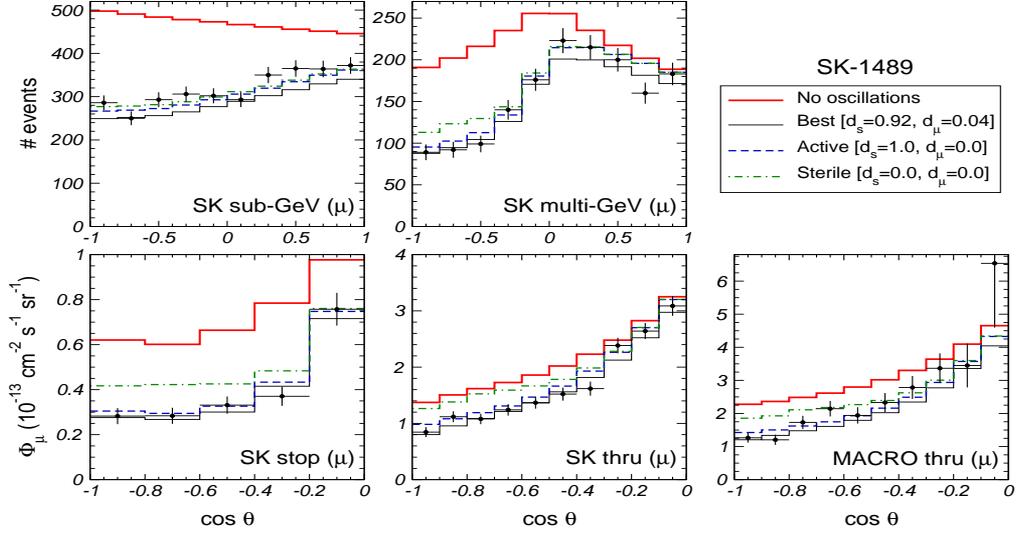}
    \caption{\label{fig:atm-zenith}%
      Zenith angle dependence of the $\mu$-like atmospheric neutrino
      data from Ref.~\cite{maltoni:2002ni}.  From the figure one can
      compare the predicted number of atmospheric neutrino events for
      best--fit, pure--active and pure--sterile oscillations and no
      oscillations.}
\end{figure}

The observed zenith angle distributions of atmospheric neutrino events
and those expected in the Standard Model and within various
oscillation hypothesis are given in Fig.~\ref{fig:atm-zenith}.
Clearly, active neutrino oscillations describe the data very well
indeed. In contrast, the no-oscillations hypothesis is clearly ruled
out. On the other hand, conversions to sterile neutrinos lead to an
excess of events for neutrinos crossing the core of the Earth, in all
the data samples except sub-GeV.

\begin{figure*}[tbph] \centering
  \includegraphics[width=0.5\textwidth,height=4.5cm]{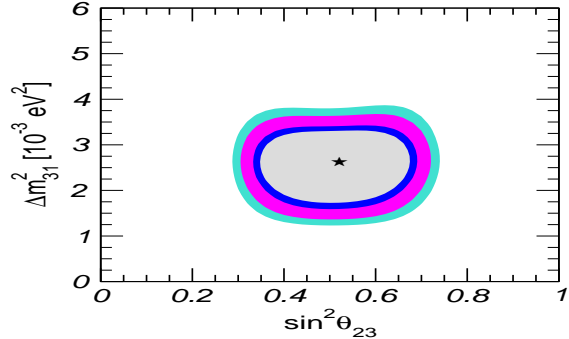}
\vspace*{-0mm}
  \caption{ \label{fig:atm-osc-par}%
    Allowed $\sin^2\theta_\Atm$ and $\Dma$ values at 90\%, 95\%, 99\%
    \CL and 3$\sigma$ corresponding to the latest atmospheric neutrino
    data, from~\cite{maltoni:2002ni}. }  \vspace*{-0mm}
\end{figure*}

\begin{figure}
\includegraphics[width=.9\linewidth,height=3.5cm]{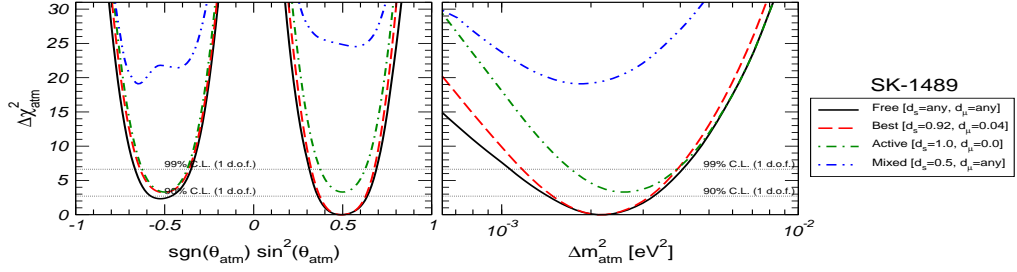}
\caption{\label{fig:chi-atm}%
  $\Dcq_\Atm$ as a function of $\Dma$, $\sin^2\theta_\Atm$, $d_s$ and
  $d_\mu$, optimizing over the undisplayed
  parameters~\cite{maltoni:2002ni}.}
\end{figure}

The parameters $\theta_\Atm$ and $\Dma$ determined from the fit are
summarized in Figs.~\ref{fig:atm-osc-par} and \ref{fig:chi-atm}.
The latter considers several cases: arbitrary $d_s$ and $d_\mu$,
best--fit $d_s$ and $d_\mu$, and pure active and mixed active--sterile
neutrino oscillations. The meaning of these $d$--parameters, not taken
into account by the Super-K collaboration, is discussed in
\cite{maltoni:2001bc}. Their existence is understood from the
structure of the 4-neutrino lepton mixing
matrix~\cite{schechter:1980gr}.
In conclusion, one finds that the improved fit of the atmospheric data
lead to a more stringent constraint on the sterile component in
atmospheric oscillations: if the \nm is restricted to the atmospheric
mass states only a sterile admixture of 16\% is allowed at 99\%\CL\,
while a bound of 35\% is obtained in the unconstrained case. Pure
sterile oscillations are disfavored with a $\Dcq = 34.6$ compared to
the pure active case.

\subsection{Solar}

In the presence of light sterile neutrinos the electron neutrino
produced in the sun converts to $\nu_x$ (a combination of $\nu_\mu$
and $\nu_\tau$) and a sterile neutrino $\nu_s \:$: $ \nu_e \to
\sqrt{1-\eta_s}\, \nu_x + \sqrt{\eta_s}\, \nu_s$.  The solar neutrino
data are fit with three parameters: $\Dms$, $\theta_\Sol$ and the
parameter $0\le \eta_s \le 1$ describing the sterile neutrino
fraction.

\begin{figure}
\includegraphics[height=4.5cm,width=0.6\linewidth]{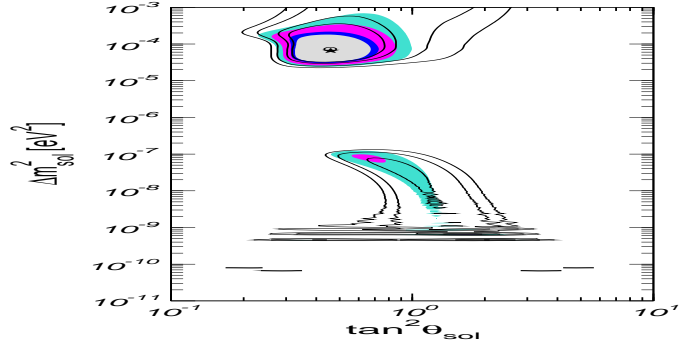}
  \caption{ \label{fig:sol-osc-par}%
    Allowed $\tan^2\theta_\Sol$ and $\Dms$ regions for $\eta_s = 0$
    (active oscillations).  The lines and shaded regions correspond to
    the \snocc\ and \snotot analyses, respectively, as defined in
    Ref.~\cite{maltoni:2002ni}.  The 90\%, 95\%, 99\% \CL and
    3$\sigma$ contours are for 3 \dof.  }
\end{figure}

In Fig.~\ref{fig:sol-osc-par} we display the regions of solar neutrino
oscillation parameters for 3 \dof\ with respect to the global minimum,
for the standard case of active oscillations, $\eta_s = 0$.
Notice that the SNO NC, spectral, and day-night data lead to an
improved determination of the oscillation parameters: the shaded
regions after their inclusion are much smaller than the hollow regions
delimited by the corresponding \snocc\ confidence contours. This is
especially important in closing the LMA-MSW region from above: values
of $\Dms > 10^{-3}~\eVq$ appear only at more than $3\sigma$. Previous
solar data on their own could not close the LMA-MSW region, only the
inclusion of reactor data~\cite{apollonio:1999ae} probed the upper
part of the LMA-MSW region~\cite{gonzalez-garcia:2000sq}. The complete
\snotot\ information is also important in excluding maximal solar
mixing in the LMA-MSW region at $3\sigma$.

\begin{figure*}[ht] \centering
\includegraphics[height=3.5cm,width=0.98\textwidth]{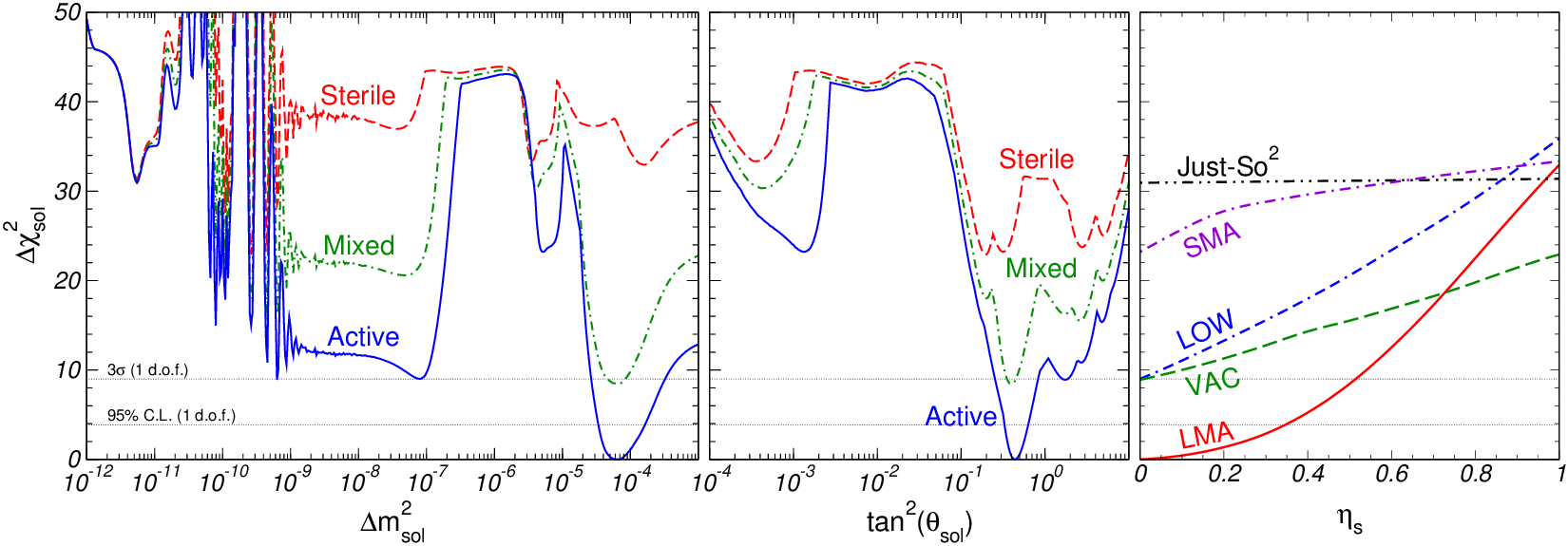}
\vspace*{-0mm}
  \caption{ \label{fig:chi-sol}%
    $\Delta\chi^2_\Sol$ versus $\Dms$, $\tan^2\theta_\Sol$, and $0
    \leq \eta_s \leq 1$ from global \snotot\ sample in
    Ref.~\cite{maltoni:2002ni}.}  \vspace*{-0mm}
\end{figure*}

Fig.~\ref{fig:chi-sol} gives the profiles of $\Delta\chi^2_\Sol$ as a
function of $\Dms$ (left), $\tan^2\theta_\Sol$ (middle) as well as
$\eta_s$ (right), by minimizing with respect to the undisplayed
oscillation parameters.
In the left and middle panels the solid, dashed and dot-dashed lines
correspond to $\eta_s = 0$, $\eta_s = 1$ and $\eta_s = 0.5$,
respectively.

\subsection{Solar + KamLAND}

The KamLAND collaboration has detected reactor neutrinos at the
Kamiokande site coming from nuclear plants at distances 80-350 km
away, with an average baseline of about 180 km, long enough to test
the LMA-MSW region~\cite{eguchi:2002dm}. The target for the
$\overline{\nu}_e$ flux is a spherical transparent balloon filled with
1000 tons of non-doped liquid scintillator, and the antineutrinos are
detected via the inverse neutron $\beta$-decay process
$\overline{\nu}_e + p \to e^+ + n$. KamLAND has for the first time
observed the disappearance of neutrinos produced in a power reactor
during their flight over such distances. The observed--to--expected
event number ratio is $0.611\pm 0.085 {\rm (stat)} \pm 0.041 {\rm
  (syst)} $ for $\bar{\nu}_e$ energies $>$ 3.4 MeV, giving the first
terrestrial confirmation of the solar neutrino anomaly with
man-produced neutrinos.

\begin{figure}
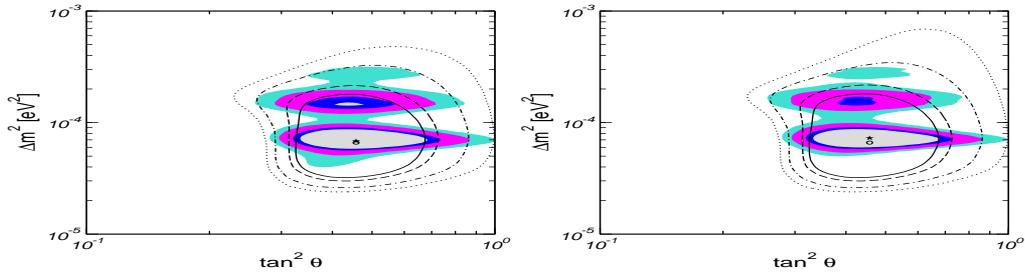

\includegraphics[height=3.5cm,width=.45\linewidth]{fig.region.eps}
\includegraphics[height=3.5cm,width=.45\linewidth]{fig.region-pois.eps}
 \caption{\label{fig:region}%
   Allowed 90\%, 95\%, 99\% and 99.73\% \CL\ regions (2~\dof) from the
   global analysis of solar, Chooz and KamLAND
   data~\cite{maltoni:2002aw}. The hollow lines do not include
   KamLAND. Left is for Gaussian, right is for Poisson statistics. The
   star (dot) is the best fit point for combined (solar+Chooz only)
   analysis.}
  \end{figure}

\begin{figure}
  \includegraphics[height=3.2cm,width=.9\linewidth]{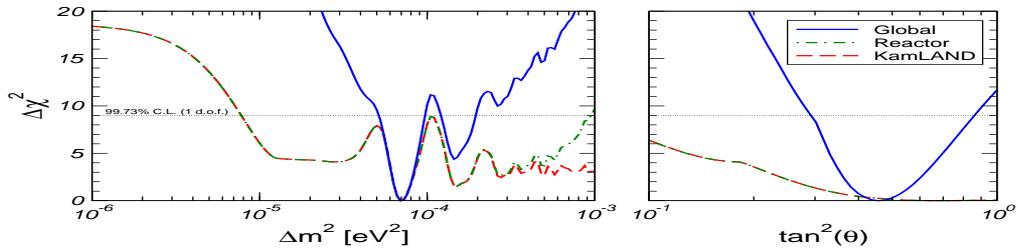}
 \caption{\label{fig:chisq}%
   $\Dcq$ profiles versus $\Dms$ and $\tan^2 \theta$ from
   \cite{maltoni:2002aw}. The (red) dashed line refers to KamLAND
   alone. The (green) dot-dashed line corresponds to the full reactor
   data sample, including both KamLAND and Chooz.  The (blue) solid
   line refers to the global analysis of the complete solar and
   reactor data.}
\end{figure}

The impact of combining the first 145.1 days of KamLAND data with the
full sample of solar neutrino data on the determination of neutrino
oscillation parameters, is shown in Figs.~\ref{fig:region}
and~\ref{fig:chisq}, from~\cite{maltoni:2002aw}. One finds that
non-oscillation solutions~\cite{barranco:2002te,guzzo:2001mi} are now
rejected at more than 3~$\sigma$, while non-LMA-MSW oscillations are
excluded at more than 4~$\sigma$.  Furthermore, the new data have a
strong impact in narrowing down the allowed range of $\Dms$ inside the
LMA-MSW region.  In contrast, the new data have little impact on the
location of the best fit point.  In particular the solar neutrino
mixing remains significantly non-maximal (3~$\sigma$).

The result for the fit shows a clear preference of the data for the
pure active LMA-MSW solution of the solar neutrino problem, with the
LOW, VAC, SMA-MSW and Just-So$^2$ solutions disfavored by a $\Dcq =
22, 22, 36, 44$, respectively.  The global solar data constrains the
admixture of a sterile neutrino to be less than 43\% at 99\% \CL.


\subsection{Robustness of the oscillation parameter determination}

How robust is the determination of solar and atmospheric neutrino
oscillation parameters, taking into account the possible existence of
other non-standard neutrino properties?
Many models of neutrino mass are acompanied by potentially sizable
non-standard neutrino interactions, which may be flavour-changing (FC)
or non-universal (NU), arising either from
gauge~\cite{schechter:1980gr} or Yukawa
interactions~\cite{hall:1986dx}.  These may affect neutrino
propagation properties in matter even in the massless
limit~\cite{valle:1987gv,guzzo:1991hi,nunokawa:1996tg}.

\begin{figure}[t]
    \includegraphics[width=0.6\linewidth,height=6cm]{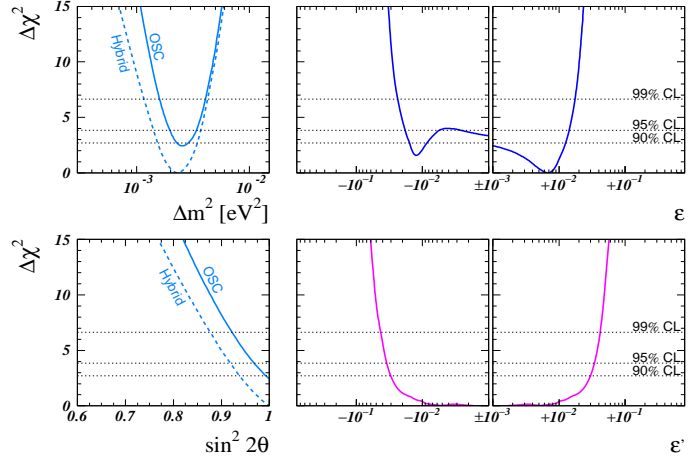}
    \caption{ \label{fig:new_chisq} %
      Behaviour of $\Delta\chi^2$ for the new Super--K and MACRO data,
      as a function of the oscillation parameters $\Delta m^2$ and
      $\theta$ (left panels), and of the NSI parameters $\Eps$ and
      $\Epp$ (right panels)~\cite{fornengo:2001pm}. In the left panels
      both the pure oscillations case and the hybrid OSC + NSI
      mechanism are given. Optimizing over undisplayed parameters is
      performed in all cases.}
\end{figure}

In Ref.~\cite{fornengo:2001pm} the atmospheric neutrino anomaly has
been reconsidered in light of the latest data from Super-K contained
events and from Super-K and MACRO up-going muons.  Neutrino evolution
was described in terms of non-standard neutrino-matter interactions
(NSI) and conventional \nm to \nt oscillations (OSC). The statistical
analysis of the data shows that a pure NSI mechanism is now ruled out
at 99\%, while standard \nm to \nt oscillations provide an excelent
description of the anomaly.  Limits were derived on FC and NU neutrino
interactions, as illustrated in Fig.~\ref{fig:new_chisq}.
One sees that the off-diagonal flavour-changing neutrino parameter
$\epsilon$ and the diagonal non-universality neutrino parameter
$\epsilon'$ are confined to $-0.03 < \epsilon < 0.02$ and $|\epsilon'|
< 0.05$ at 99.73\% CL. These limits are model independent as they are
obtained from pure neutrino-physics processes. The stability of the
neutrino oscillation solution to the atmospheric neutrino anomaly
against the presence of NSI establishes the robustness of the
near-maximal atmospheric mixing and massive-neutrino hypothesis.  The
current sensitivity of atmospheric neutrino experiments to the
existence of neutrino flavor changing NSI can be further improved with
future neutrino factory experiments, especially for higher
energies~\cite{huber:2001zw}.

It has been shown that non-standard neutrino interactions give a very
good interpretation of current solar neutrino data consistent with the
oscillation description of the atmospheric neutrino
data~\cite{guzzo:2001mi}. Such solutions, however, although preferred
by the solar data, are ruled out by the first results of the KamLAND
reactor experiment, at more than 3~$\sigma$~\cite{guzzo:2001mi}.
Therefore NSI can not be the leading explanation of the solar neutrino
anomaly.  Likewise, one can investigate how robust is the
determination of solar oscillation parameters, taking into account the
presence of NSI.

\begin{figure}[htb]
 \includegraphics[width=0.8\textwidth,height=3.2cm]{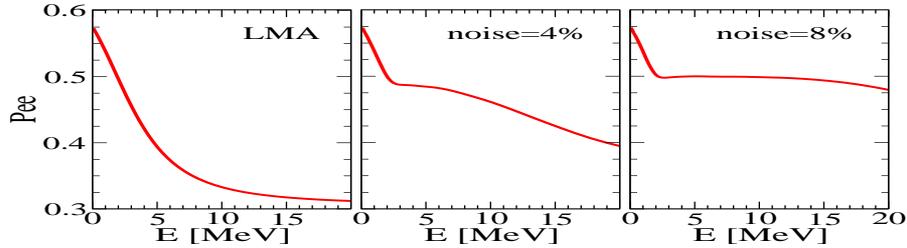}
    \caption{Effect of random matter density perturbations on the
      LMA-MSW solution. The left panel is for quiet Sun, middle and
      right panels correspond to $\xi=4 \%$ and $\xi=8 \%$,
      respectively, from \cite{burgess:2002we}} 
    \label{fig:noisyLMA}
\end{figure}
\begin{figure}[htb!]
   \includegraphics[width=.8\textwidth,height=4cm]{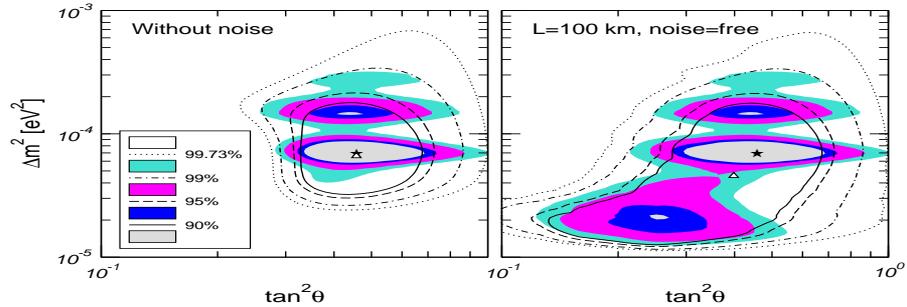}
    \caption{Allowed regions of neutrino
      oscillation parameters for the quiet Sun (left panel) and for a
      noisy Sun on a $L_0=100$ km spatial scale with arbitrary density
      noise magnitude $\xi$~\cite{burgess:2002we}.  The lines and
      shaded regions correspond to the analyses of solar and
      solar+KamLAND data respectively.}
    \label{fig.osc-region.eps}
  \end{figure}

We now tuen to the robustness of solar oscillations parameter
determination taking into account uncertainties in solar physics.  One
possibility is to consider random solar matter
density~\cite{balantekin:1996pp}. It has been argued that a resonance
between helioseismic and Alfvén waves might provide a physical
mechanism for generating these fluctuations~\cite{Burgess:2003fj}.
They can have an important effect upon neutrino conversion in
matter~\cite{nunokawa:1996qu,bamert:1998jj}, as shown in
Figs.~\ref{fig:noisyLMA}.  and~\ref{fig.osc-region.eps}.
From the latter one sees that the determination of neutrino
oscillation parameters from a combined fit of KamLAND and solar data
depends strongly on the magnitude of solar density fluctuations.
  Given the current neutrino oscillation parameters, the results of
  KamLAND imply new information on fluctuations in the solar
  environment on scales to which standard helioseismic tests are
  largely insensitive. The sensitivity of present solar neutrino +
  KamLAND data to the solar noise parameter $\xi$ as a function of the
  correlation length $L_0$, when neutrino oscillation parameters are
  varied inside the present LMA region has been given in
  \cite{burgess:2002we}.

\subsection{LSND}

The LSND experiment~\cite{aguilar:2001ty}, which sees evidence for
$\bar{\nu}_e$ appearance in a $\bar{\nu}_\mu$ beam, is suggestive of
oscillations on a scale much higher than those indicated by solar and
atmospheric data. All other experiments such as the short baseline
disappearance experiments Bugey~\cite{declais:1995su} and
CDHS~\cite{dydak:1984zq}, as well as the KARMEN neutrino
experiment~\cite{armbruster:2002mp} report no evidence for
oscillations.

\begin{figure}
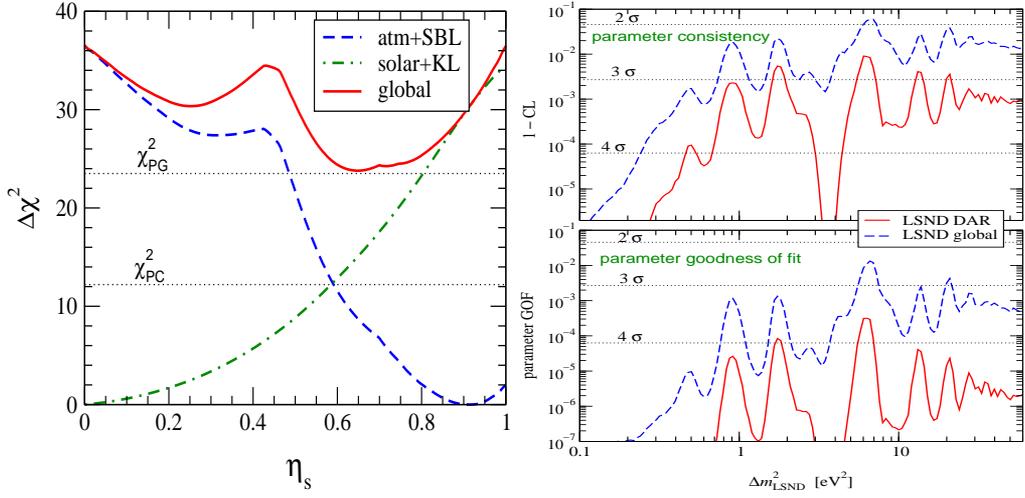

\includegraphics[width=.45\linewidth,height=6.5cm]{etas_sol-kl+atm.eps}
\includegraphics[width=.45\linewidth,height=6.5cm]{3+1.eps}
 \caption{$\Delta\chi^2_\Sol$, $\Delta\chi^2_{\Atm+\Sbl+\Kl}$
   and $\bar\chi^2_\mathrm{global}$ as a function of $\eta_s$ in (2+2)
   oscillation schemes (left). The right panel show how (3+1) schemes
   are still acceptable at 3~$\sigma$.
   From~Ref.~\cite{maltoni:2002xd}}
  \label{fig:sol+atm+kl}
\end{figure}

Prompted by recent improved solar and atmospheric data,
ref.~\cite{maltoni:2002xd} has re-analysed the four-neutrino
description of all current neutrino oscillation data, including the
LSND evidence. The higher degree of rejection for non-active solar and
atmospheric oscillation solutions implied by the SNO neutral current
result, and by the latest 1489-day Super-K atmospheric neutrino data,
allows one to rule out (2+2) oscillation schemes proposed to reconcile
LSND with the rest of current oscillation data.  Using an improved
goodness of fit (gof) method especially sensitive to the combination
of data sets one obtains a gof of only 1.6 $\times 10^{-6}$ for (2+2)
schemes.  This is illustrated by the left panel in
Fig.~\ref{fig:sol+atm+kl}.
Also shown in the left panel of Fig.~\ref{fig:sol+atm+kl} are the
values $\chi^2_\mathrm{PC}$ and $\chi^2_\mathrm{PG}$ relevant for
parameter consistency and parameter \gof, respectively. The right
panel displays the compatibility of LSND with solar+atmospheric+NEV
data in (3+1) schemes. In the upper panel we show the \CL\ of the
parameter consistency whereas in the lower panel we show the parameter
\gof\ for fixed values of $\Dml$. The analysis is performed both for
the global \cite{aguilar:2001ty} and for the DAR \cite{church:2002tc}
LSND data samples.

In summary one finds that the strong preference of oscillations into
active neutrinos implied by solar+KamLAND, as well as atmospheric
neutrino data, rules out (2+2) mass schemes, whereas (3+1) schemes are
strongly disfavoured, but not ruled out, by short-baseline
experiments.

\begin{figure}
\includegraphics[width=.6\linewidth,height=4.7cm]{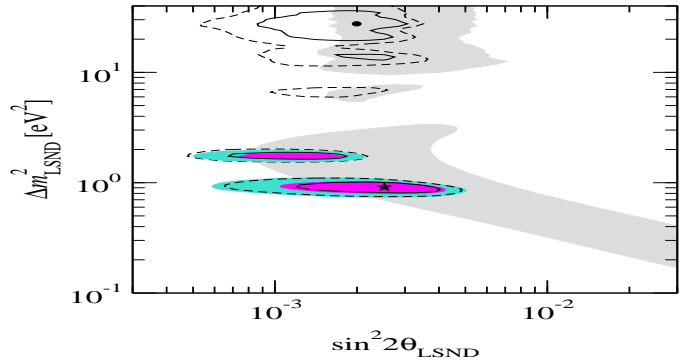}
  \caption{\label{4-nu-lab-cosmo} Allowed regions at 90 and 99 \%\CL\
    for 3+1 schemes with (clored) and without (solid and dashed lines)
    information coming from cosmology. The grey region is the 99\% \CL\
    LSND region~\cite{maltoni:2003yr}}
\end{figure}

There are, in addition, recent data from cosmology, including CMB data
from WMAP~\cite{spergel:2003cb} and data from 2dFGRS large scale
structure surveys~\cite{peacock:2001gs} that can be used to further
constrain 4-neutrino schemes~\cite{elgaroy:2003yh,elgaroy:2002bi}.
The result of such analysis, based on the data
of~\cite{hannestad:2003xv} is illustrated in
Fig.~\ref{4-nu-lab-cosmo}.
Thus one sees that cosmology cuts the large $\Dml$ values. In summary,
one finds that even (3+1) schemes are strongly disfavoured by current
data, bringing the LSND hint to a more puzzling status that led
people to suggest solutions as radical as the violation of
CPT~\cite{pakvasa:2003zv}.

\section{Three-neutrino parameters}

\begin{figure}
\includegraphics[height=5cm,width=.8\linewidth]{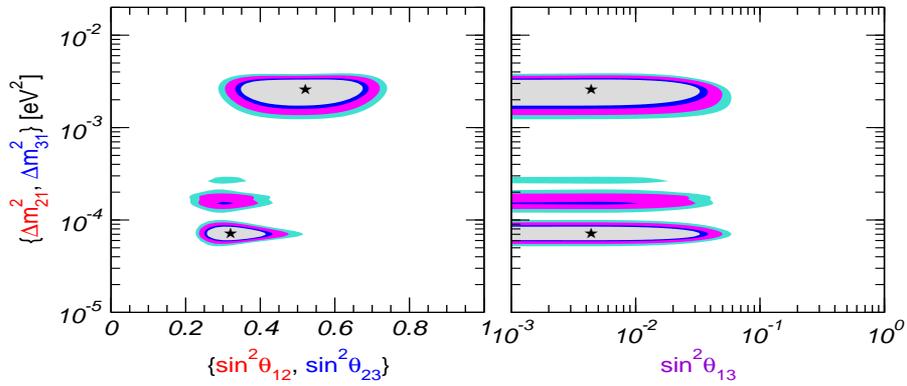}
   \caption{\label{3-nu-regions} 90, 95, 99\% \CL\ and 3~$\sigma$ 
     regions of solar and atmospheric oscillation paramaters versus
     the corresponding mixing parameters (left) and versus
     $\sin^2\theta_{13}$ from \cite{maltoni:2003}}
 \end{figure}

\begin{figure}
\includegraphics[height=3cm,width=.9\linewidth]{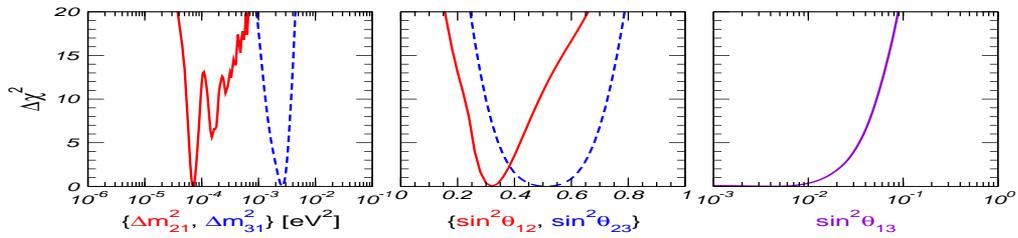}
  \caption{\label{3-nu-chi2} Determinining the five 3-neutrino oscillation
    parameters, from \cite{maltoni:2003}}
\end{figure}

\begin{figure}[htbp]
\includegraphics[width=4cm,height=3.5cm]{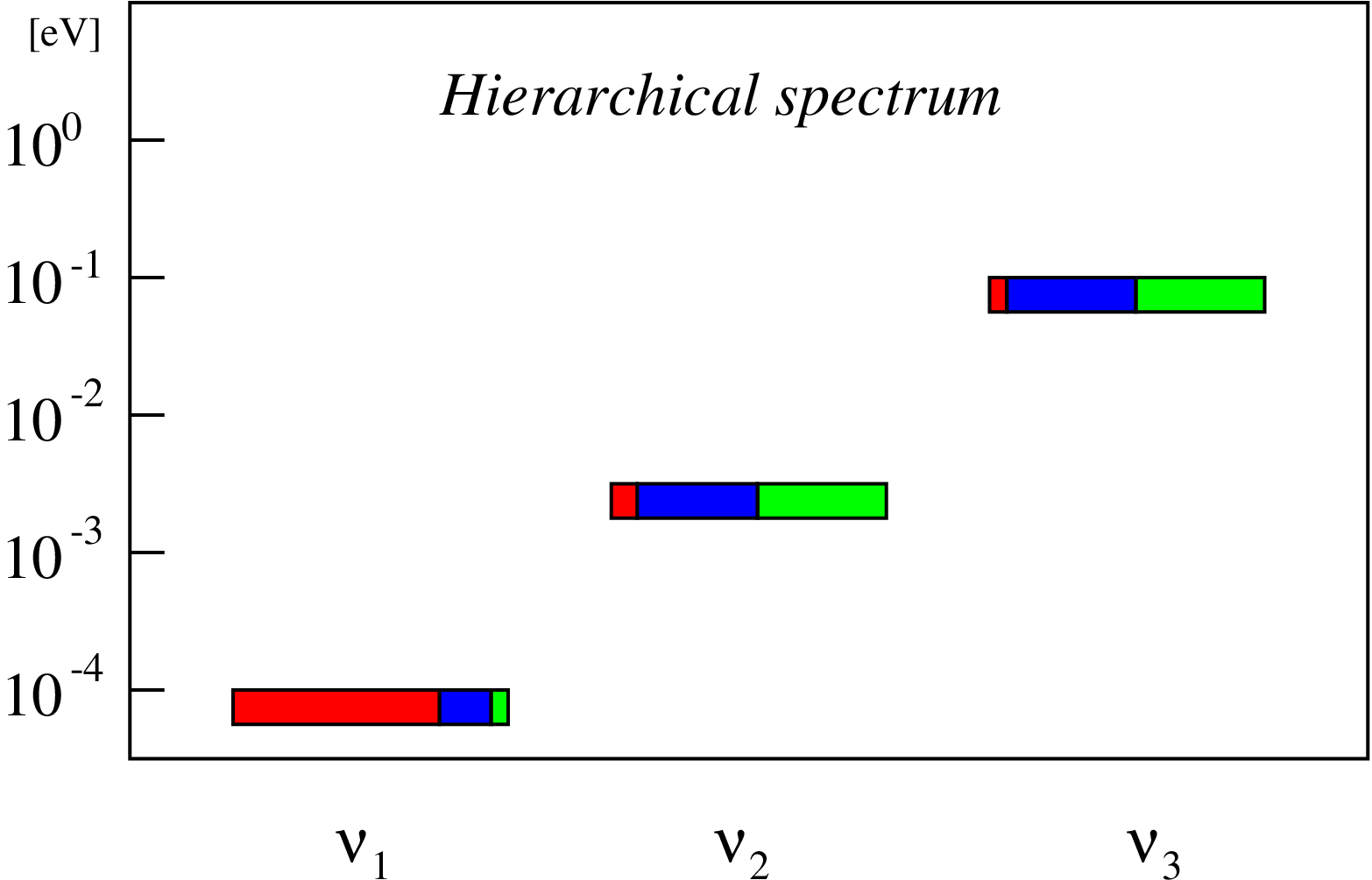}
\includegraphics[width=4cm,height=3.5cm]{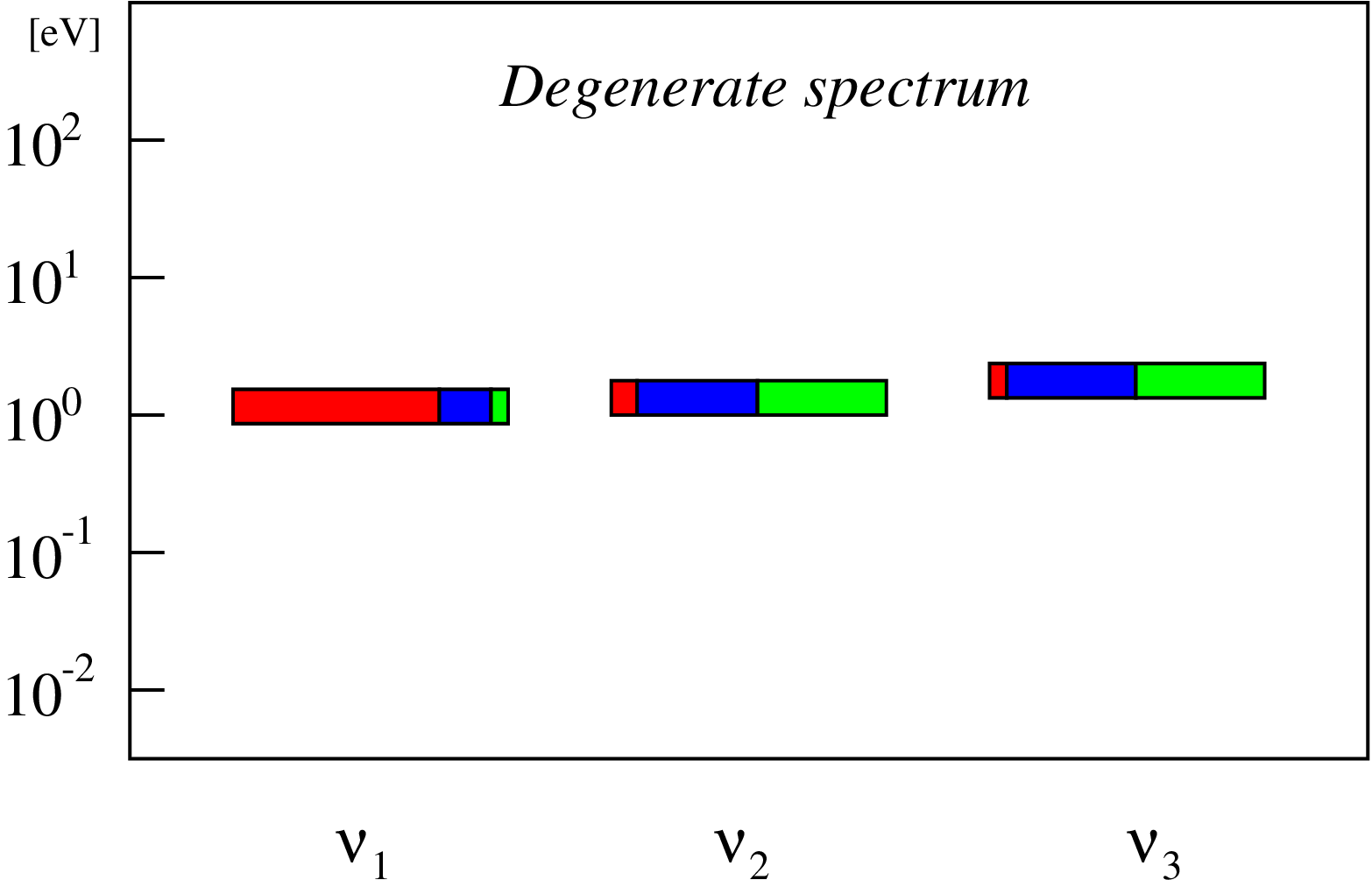}
\caption{\label{nu-spectra} Hierarchical (left) and quasi-degenerate
    neutrino spectra (right). }
\end{figure}

Given the stringent bounds on four-neutrino schemes derived from
current global fits it is relevant to analyse in more detail the
constraints on three-neutrino parameters, in particular on
$\theta_{13}$, so far neglected. This has been done in detail in
Ref.~\cite{gonzalez-garcia:2000sq}. Here we summarize the update to be
presented in \cite{maltoni:2003} including the most recent data, such
as the K2K data which has recently observed positive indications of
neutrino oscillation in a 250 km long-baseline
setup~\cite{ahn:2002up}. The projections of the five-dimensional
parameter space are displayed in Fig.~\ref{3-nu-regions}.
 The goodness of the determination of the five 3-neutrino oscillation
 parameters is illustrated in Fig.~\ref{3-nu-chi2}.
Both hierarchical and quasi-degenerate neutrino mass spectra,
illustrated in Fig.~\ref{nu-spectra}, are compatible with current
data.  \vskip .3cm

\subsection{Neutrino factories and the angle $\theta_{13}$}

Neutrino factories aim at probing the lepton mixing angle
$\theta_{13}$ with much better sensitivity than possible at present,
and thus open the door to the possibility of leptonic CP
violation~\cite{apollonio:2002en,freund:2001ui}.  We have already
discussed both the hierarchical nature of neutrino of mass splittings
indicated by the observed solar and atmospheric neutrino anomalies, as
well as the stringent bound on $\theta_{13}$ that follows from reactor
experiments Chooz and Palo Verde. We also mentioned that the leptonic
CP violation associated to the standard Dirac phase present in the
simplest three-neutrino system disappears as two neutrinos become
degenerate and/or as $\theta_{13} \to 0$~\cite{schechter:1980bn}.  As
a result, although the large mixing associated to LMA-MSW certainly
helps, direct leptonic CP violation tests in oscillation experiments
will be a very demanding task for neutrino factories.

Refs.~\cite{huber:2001de,huber:2002bi} considered the impact of
non-standard interactions of neutrinos on the determination of
neutrino mixing parameters at a neutrino factory using
$\pnu{e}\to\pnu{\mu}$ ``golden channels'' for the measurement of
$\theta_{13}$. One finds that even a small residual NSI leads to a
drastic loss in sensitivity in $\theta_{13}$, of up to two orders of
magnitude~\footnote{This can be somewhat overcome if two baselines are
  combined.}.

 \begin{figure}[htb!]
    \includegraphics[width=0.60\textwidth,height=4.5cm]{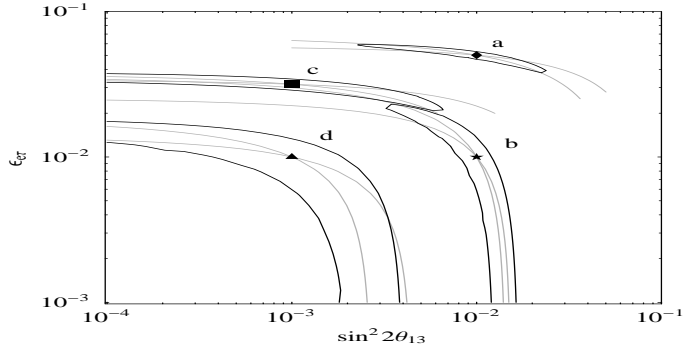}
    \caption{99\% \CL\ allowed regions (black lines) in 
      $\sin^2 2\theta_{13}$--$\epsilon_{e\tau}$ for different input
      values indicated by the points, at a baseline of
      $3\,000\,\mathrm{km}$, from \cite{huber:2001de}. Lines of
      constant event rates are displayed in grey.}
    \label{fig:fits}
\end{figure}
Therefore the design of a neutrino factory should include a near-site
detector capable of monitoring the possible presence of non-standard
interactions. For more details on the confusion between NSI and
oscillations in the e-tau channel see ~\cite{huber:2002bi}.

\section{Absolute neutrino mass scale}
  
Oscillations are sensitive only to neutrino mass splittings and mixing
angles, not to the absolute scale of neutrino mass.  Tritium end-point
~\cite{hagiwara:2002fs} and \nbb
experiments~\cite{klapdor-kleingrothaus:1999hk,morales:1998hu} may
determine the absolute scale of neutrino
mass~\cite{klapdor-kleingrothaus:2000gr}.

\begin{figure}[htbp]
  \centering
 \includegraphics[width=.45\linewidth,height=4cm]{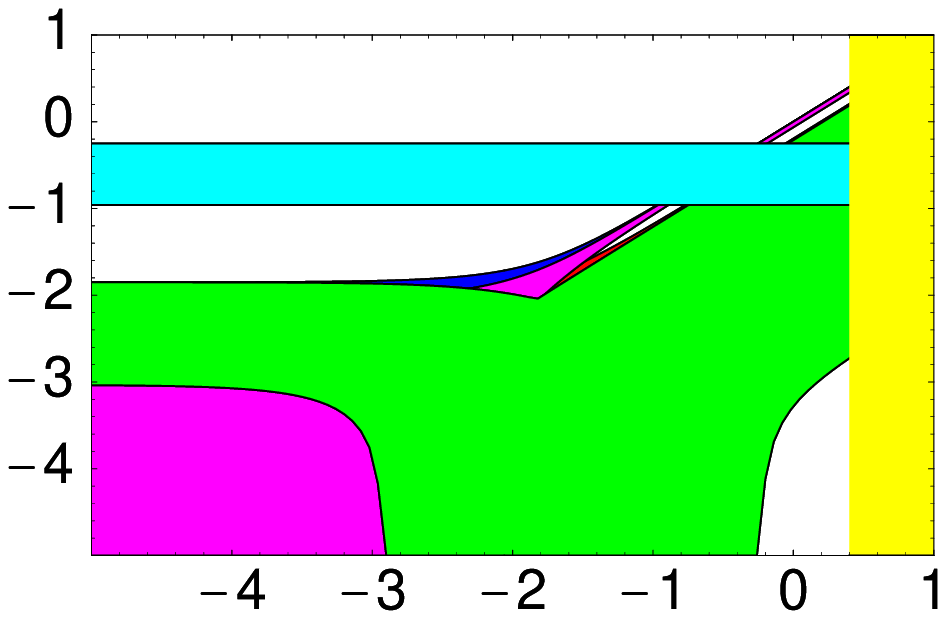}
\includegraphics[width=.45\linewidth,height=4cm]{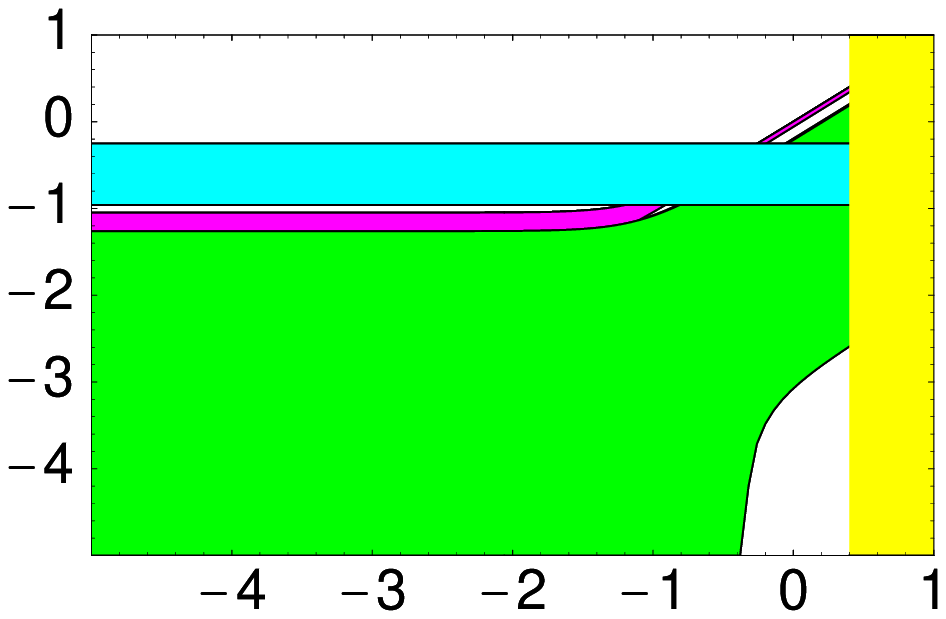}
  \caption{\label{nbb}Log~$\meff$/eV
    versus Log~$m_1$/eV, \nbb (horizontal band) and tritium
    sensitivities (vertical)}
  \label{fig:nbb}
\end{figure}

In contrast to the 2-neutrino mode, neutrinoless double beta
decay~\cite{morales:1998hu,faessler:2001kd} violates L by 2 units. It
is expected to occur due to the exchange of massive neutrinos (mass
mechanism), provided they are Majorana particles. The phase space
advantage of \nbb opens some hope of overcoming the suppression due to
the L-violating Majorana neutrino propagator~\cite{schechter:1981hw}.
Now that neutrino masses have been established, one expects a
non-vanishing \nbb decay rate.  The amplitude is proportional to the
parameter $ \meff = \sum_j K_{ej}^2 m_j $ which can be given as,
$$
 \meff     =  c_{12}^2 c_{13}^2  m_1
           +  s_{12}^2 c_{13}^2 e^{i \alpha}  m_2 + 
              s_{13}^2 e^{i \beta}        m_3
              $$
              in terms of three masses $m_i$, two angles
              $\theta_{12}$ and $\theta_{13}$, and two CP violating
              phases: $\alpha, \beta$.  Using current neutrino
              oscillation data one can display the attainable $\meff$
              values for a normal (left) versus an inverse hierarchy
              (right), as shown in Fig.~\ref{fig:nbb}. Different
              shades (colors) correspond to different CP sign
              combinations among the three neutrinos.
One sees from Fig.~\ref{fig:nbb} that \nbb is too small if the
neutrinos obey a normal mass hierarchy ($\meff \lsim 0.01$ eV at left
part of left panel). In contrast, it is enhanced by about an order of
magnitude if the hierarchy is inverted ($\meff \lsim 0.1$ eV at left
part of right panel), or in the quasi-degenerate limit ($\meff \sim 1$
eV right parts of either panel in Fig.~\ref{fig:nbb}). Progress in
this field will come about through the improvement of the current
upper limit for $\meff \le 0.3 $ eV in a new generation of \nbb
experiments, such as
GENIUS~\cite{klapdor-kleingrothaus:1999hk,morales:1998hu}, an
improvement in the accuracy of nuclear matrix elements (currently a
with factor $\sim 2$)~\cite{faessler:2001kd,elliott:2002xe}, as well
as an improvement in the current upper limit from tritium experiments:
$m_1 \le 2.5 $ eV in experiments such as
KATRIN~\cite{osipowicz:2001sq}.

\begin{figure}[htbp]
  \centering
\includegraphics[width=6cm,height=3.5cm]{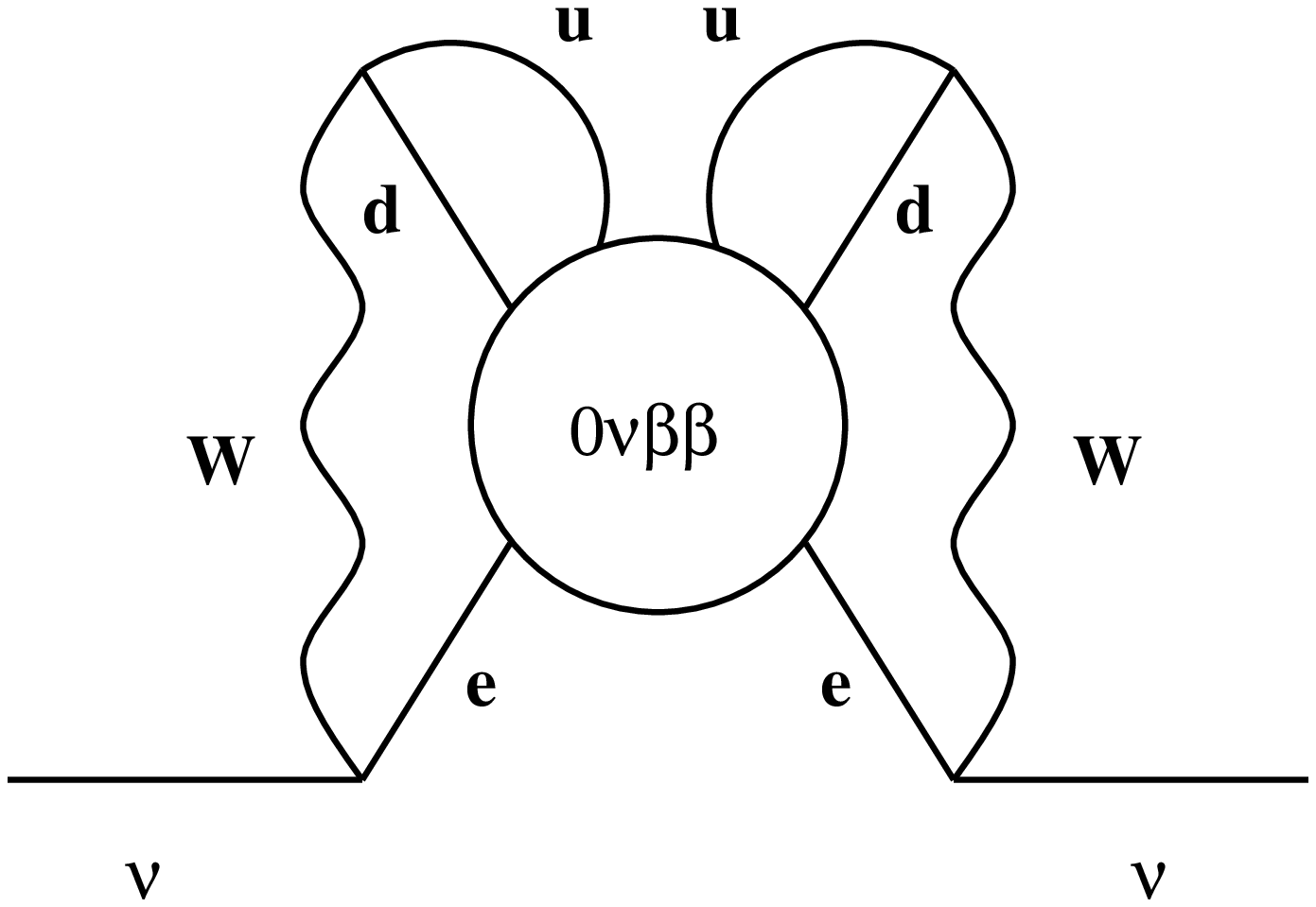}  
  \caption{ \label{fig:bbox} The black-box \nbb argument, 
from \cite{schechter:1982bd}.}
\end{figure}

The importance of \nbb in deciding the nature of neutrinos goes far
beyond the details of the mass mechanism.  The connection between the
two is given by the ``black-box theorem'' which states what, in a
gauge theory, whatever the mechanism for inducing \nbb is, it is bound
to also yield a Majorana neutrino mass at some level, and vice-versa,
as illustrated by fig.  \ref{fig:bbox} \cite{schechter:1982bd}.

\section{Neutrino theories}

\begin{figure}
\includegraphics[width=.5\linewidth,height=3.5cm]{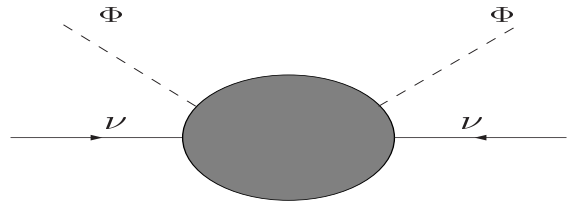}
  \caption{ \label{fig:d5} Dimension
    5 operator for neutrino mass.}
\end{figure}

\begin{figure}
\includegraphics[width=.6\linewidth,height=2.8cm]{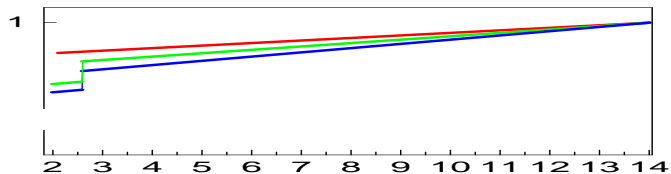}
  \caption{\label{fig:nu-unif} Neutrino mass in eV versus
    Log~$M_X$/GeV, where $M_X$ is the neutrino unification scale }
\end{figure}

The simplest way to generate neutrino masses is via the
dimension--five operator of Fig.~\ref{fig:d5}~\cite{weinberg:1980mh}.
Such may arise from gravity itself, though in this case the masses
generated are much smaller than indicated by current solar and
atmospheric experiments.
Thus one needs to appeal to physics at a lower scale. The seesaw with
a scale in the unification range can do the job~\cite{ross:2002fb}. An
extreme view of this approach is the idea of neutrino unification,
advocated in~\cite{chankowski:2000fp}~\footnote{Only the CP conserving
  variant of the model in \cite{chankowski:2000fp} is ruled out by
  current solar data.} and \cite{babu:2002dz}, and illustrated in
Fig.~\ref{fig:nu-unif}.
In this picture, neutrino masses behave like the gauge coupling
constants of the Standard Model~\cite{langacker:1995fk}, which merge
at high energies as they are evolved via the renormalization group,
due to the presence of
supersymmetry~\cite{casas:1999ac,frigerio:2002in,antusch:2002rr}.
Such a simple theoretical ansatz provides the most natural framework
for having a neutrino mass scale potentially observable in tritium and
\nbb experiments, as well as cosmology, leading to the so-called
quasi-degenerate neutrino spectrum (right panel of
Fig.~\ref{nu-spectra}.

In the variant of this idea proposed in \cite{babu:2002dz} the
discrete non-Abelian symmetry $A_4$, valid at some high-energy scale,
is responsible for the degenerate neutrino masses, without spoiling
the hierarchy of charged-lepton masses.  Mass splittings and mixing
angles are induced radiatively in the context of softly broken
supersymmetry.  The model predicts that the atmospheric angle is
necessarily maximal and that either the mixing parameter $\theta_{13}$
is zero or pure imaginary, leading to maximal CP violation in neutrino
oscillations~\cite{grimus:2003yn}.  The solar angle is unpredicted,
but can be large.  The quark mixing matrix is also calculable in a
similar way. Large lepton mixing angles are compatible with the
smallness of the quark mixing angles because only neutrinos are
degenerate.  Neutrinoless double beta decay and flavor violating tau
decays such as $\tau \to \mu \gamma$ should be in the experimentally
accessible range.

Low energy supersymmetry as the origin of neutrino
mass~\cite{diaz:2003as,hirsch:2000ef,romao:1999up} provides a viable
alternative to the seesaw.  In this case Weinberg's operator arises
from weak--scale physics~\cite{diaz:2003as,hirsch:2000ef,romao:1999up}
in theories with spontaneous breaking of R
parity~\cite{ross:1985yg,ellis:1985gi,masiero:1990uj}.  These lead
effectively to bilinear violation~\cite{diaz:1998xc}.  Neutrino mixing
angles can be probed at accelerator experiments, which have therefore
the potential to falsify the model.  Alternative low energy mechanisms
for neutrino mass generation are the models of Babu~\cite{babu:1988ki}
and Zee~\cite{zee:1980ai} and variants thereof.

All in all, there is no clear ``road map'' for the ultimate theory of
neutrinos, since we can not predict neutrino properties from first
principles: we do not know the underlying scale nor the mechanism
involved. Last, but not least, we lack a fundamental theory of
flavour.  The neutrino mass tale is far from told, despite the
extraordinary revolution brought about by the experimental discovery
of neutrino mass.


\begin{theacknowledgments}
  I am thankful to Joe Schechter for his insights and friendship.
  This work was supported by a Humboldt Research Award (J.  C.  Mutis
  Prize), by Spanish grant BFM2002-00345, by the European Commission
  RTN grant HPRN-CT-2000-00148 and the European Science Foundation
  Neutrino Astrophysics Network. I thank Martin Hirsch for discussions
  on double beta decay, M Maltoni and M A Tórtola for advancing
  results of \cite{maltoni:2003}
\end{theacknowledgments}


\bibliographystyle{aipproc}   



\IfFileExists{\jobname.bbl}{}
 {\typeout{}
  \typeout{******************************************}
  \typeout{** Please run "bibtex \jobname" to optain}
  \typeout{** the bibliography and then re-run LaTeX}
  \typeout{** twice to fix the references!}
  \typeout{******************************************}
  \typeout{}
 }

\end{document}